\documentstyle[epsfig]{l-aa}

\newcommand{\mic}{$\mu \rm m$}
\newcommand{\gr}{$\gamma$-ray}
\newcommand{\grs}{$\gamma$-rays}
\begin{document}
 
\thesaurus{11.02.2: Mkn~501;12.04.2;13.09.2;13.07.2;13.25.2}

\title{Constraints on the Cosmic Infra-Red Background
 based on BeppoSAX and CAT spectra of Mkn~501}

\author{
J. Guy\inst{1}, C. Renault\inst{1}, F.A. Aharonian\inst{2}, M. Rivoal\inst{1}, J.-P. Tavernet\inst{1}
}
\institute{
 LPNHE, CNRS-IN2P3 Universit\'es Paris VI-VII, 4 place
 Jussieu, F-75252 Paris Cedex 05, France
 \and
 Max Planck Institut f\"ur Kernphysik,
 Postfach 103980, D-69029 Heidelberg, Germany 
}

\offprints{C. Renault (rcecile@in2p3.fr)}

\date{Received;Accepted} 
\maketitle
\markboth{Guy~et~al.: Constraints on the Cosmic Infra-Red Background ...}{}

\begin{abstract} 
The TeV and X-ray data obtained by the 
imaging Cherenkov telescope CAT and X-ray satellite BeppoSAX 
during the remarkable flare of Mkn~501 in April~16,~1997
are used to constrain  the flux of the Cosmic Infrared Background (CIB)
using  different CIB  models.  
We show that a non-negligible absorption of $\gamma$-rays  
due to the CIB could take place already in the low-energy
(sub-TeV) domain of the spectrum of Mkn~501. 
This implies that the data of the low-energy threshold  CAT telescope
contain very important information about the CIB at short 
wavelengths, 0.4~$\mu$m$\leq \lambda \leq$~3.~\mic. The 
 analysis of almost simultaneous  spectroscopic 
measurements of Mkn~501 in a high state by CAT and BeppoSAX 
in the framework  of  the standard homogeneous 
 Synchrotron-Self-Compton (\mbox{SSC}) framework
model leads to the conclusion that the density of 
the near-infrared background with typical ``starlight spectrum''   
around 1~$\mu$m should be between 5 and $35 \, \rm nW \, m^{-2} \, sr^{-1}$ (99$\%$~CL), 
with most likely value around 20~$\rm nW \, m^{-2} \, sr^{-1}$. 
Also we argue that  the CAT \gr~data
alone allow rather robust upper limits on the CIB,
$\lambda F_\lambda \leq 60 \ \rm nW \, m^{-2} \, sr^{-1}$ at 1~\mic,
taking into account  that for any reasonable scenario of \gr~production 
the differential intrinsic spectrum of \grs~hardly could be flatter than 
${\rm d} N/{\rm d} E \propto E^{-1}$. 
This estimate agrees, within statistical and 
systematic uncertainties,  with recent reports about 
tentative detections of the CIB  at 2.2 and 3.5 $\mu$m 
by the Diffuse Infrared Background Experiment (DIRBE),  
as well as  with the measurements of the 
background radiation at optical wavelengths from absolute photometry. 
The high flux of CIB at $\leq$~few~\mic~
wavelengths implies a significant distortion of the 
shape of the initial (source) spectrum of $\gamma$-rays from Mkn~501
at sub-TeV energies. 
The ``reconstructed'' intrinsic $\gamma$-ray spectrum shows a
distinct  peak in the Spectral Energy Distribution (SED)
around 2~TeV with a flux by a factor of 3 higher than   
the measured flux.  The energy spectrum of gamma radiation 
from both sides of the peak has power-law behavior with 
photon index $\alpha \simeq 1.5$ below 2~TeV, and
$\alpha \simeq 2.5$ above 2~TeV. This agrees with 
predictions of \mbox{SSC} model.  
We also discuss the impact of the intergalactic absorption  effect in  
derivation of the \mbox{SSC} parameters for the jet in Mkn~501.  

\end{abstract}

\keywords{BL Lacertae objects: individual: Mkn~501; Cosmology: diffuse radiation;
Infrared: general; gamma rays: observations;
X-rays: galaxies}

\section{Introduction} 

The Cosmic Infrared Background (CIB) is contributed mostly by the 
red-shifted `stellar' and `dust'  radiation components, and therefore carries 
vital cosmological information about the epoch of galaxy formation 
and their evolution in time.  The derivation of information about CIB from
direct measurements is a hard task which requires an effective 
removal  of  heavy contamination caused by foregrounds of
different origin, in particular  by the  zodiacal light, 
stellar and interstellar emission  of our Galaxy, etc 
(Arendt et~al. 1998;   Kelsall et~al. 1998).
The conclusions of this  approach in the near-infrared domain are to a 
large extent   model-dependent because they are 
generally  based  on comprehensive  modeling of the foregrounds; 
the far-infrared background is determined with a better accuracy as 
it dominates  the foreground emission (Hauser et~al. 1998).   

Very High Energy (VHE) \gr~astronomy  
provides an independent and complementary approach
for the study of the CIB. The idea is simple, and based on the
detection of absorption features in the \gr~spectra of distant
extragalactic objects  caused by interactions  of VHE \grs~with the
CIB photons in their way from a  source to the observer  
(Nikishov 1962; Gould and Schreder  1967, Stecker et~al. 1992).    
The recent  detections of  \grs~from two BL~Lac objects, Mkn~421 and Mkn~501, 
with  spectra  extending  up to 10~TeV and beyond, 
open an  interesting path for the realization of  this exciting cosmological aspect of  
VHE gamma-ray  astronomy. Obviously,  the success of  the 
`$\gamma$-astronomical'  approach  essentially depends on 
two crucial conditions: (i) accurate \gr~spectrometry, and 
(ii) good understanding of the  intrinsic (source) spectra of TeV  $\gamma$-rays.

The observations of  Mkn~501, the  second  closest X-ray selected BL~Lac 
object  with a redshift $z \simeq 0.034$, during its extraordinary outburst in 1997  
yielded  unique data,   which initiated important  
theoretical studies of the  physical  conditions in the 
relativistic  jets of BL Lac objects  
(see {\it e.g.} Tavecchio ~et~al. 1998 (hereafter TMG);  Kirk \& Mastichiadis 1999, 
Hillas 1999; Coppi \& Aharonian 1999a, 
Bednarek \& Protheroe 1999,
Krawczynski ~et~al. 1999),  
as well as interesting efforts to set meaningful   upper limits on the 
CIB flux (see   Biller ~et~al. 1998,  Stanev \& Franceschini 1998, Barrau 1998, 
 Stecker \& De Jager 1998,  Stecker 1999,
Aharonian ~et~al. 1999,  Coppi \& Aharonian 1999b; Konopelko ~et~al. 1999). 

During the 1997 outburst, lasted  several months,   Mkn~501  
showed  dramatic variations of fluxes both  in X-rays   
(BeppoSAX:  Pian ~et~al. 1998,  RXTE: Lamer \& Wagner 1998)    and  TeV \grs~
(Whipple: Catanese ~et~al. 1997,   Samuelson ~et~al.  1998;  
HEGRA: Aharonian ~et~al.  1997, 1999; 
Telescope Array: Hayashida ~et~al. 1998;  
CAT:  Djannati-Ata\"{\i} ~et~al.  1999).   
More importantly, the high TeV fluxes allowed 
monitoring of the energy spectrum of Mkn~501 on a diurnal basis, especially
during very strong flares. On several occasions truly simultaneous observations of
Mkn~501 were available in  TeV and X-ray band  (Krawczynski ~et~al. 1999). 
A special interest presents the famous April~16,~1997 flare which was observed 
by  BeppoSAX (Pian ~et~al. 1998) and low-energy threshold  ($\sim 300 \, \rm GeV$)  
Whipple and   CAT atmospheric   Cherenkov telescopes  
(Catanese ~et~al. 1997; Djannati-Ata\"{\i} ~et~al. 1999).  
Because of very large fluxes, the energy spectrum of the flare was   
obtained   with good accuracy in broad dynamical 
ranges in both X-ray (0.1-100~keV) and TeV \gr~(0.3-10~TeV)  regimes.
Within today's  most popular, the  so-called synchrotron-self-Compton (\mbox{SSC}) model  
of non-thermal high energy radiation of Mkn~501,  the data of April~16,~1997 flare 
allow to compute the intrinsic source spectrum 
of $\gamma$-rays. This information,  coupled with almost simultaneous  
spectral measurements of $\gamma$-radiation 
by CAT,  provides a good opportunity to analyze  the intergalactic absorption  
signature in the observed VHE \gr~spectrum.

Since the optical depth  $\tau_{\gamma \gamma}$  of \grs~in the intergalactic medium
increases with energy (for any reasonable spectral shape of the CIB),   
  most stringent constraints on the CIB
come from the very energetic tail of the  \gr~ spectrum  of Mkn~501. For $E \geq$~10~TeV  
the $\gamma - \gamma$ absorption is dominated  by the `dust' component of 
CIB at far infrared (FIR)  wavelengths, $\lambda \geq$ 10~\mic.   
Therefore, it is generally believed that a deep probe of CIB at  optical and near-infrared (NIR)
 wavelengths  is contingent only upon the discovery of more distant 
({\it e.g.} with $z \geq 0.1$)  VHE \gr~sources.  In this paper, however,  we show that 
the accurate measurements of the spectrum of Mkn~501 by CAT at  
sub-TeV energies already provide meaningful upper limits on the CIB flux 
at wavelengths between 0.4 and a few \mic. 
Moreover, the analysis of the X-ray and \gr~spectra of the April~16,~1997 flare 
within the homogeneous \mbox{SSC} model  allows rather conclusive estimates
of the CIB flux at such  short wavelengths. The CIB flux
$\lambda F_{\lambda} \sim 5-35 \, \rm nW\ m^{-2} sr^{-1}$ at $\lambda \sim 1$~\mic~gives
 a reasonable slope, $\nu F_\nu \propto E^{0.5}$,
in the `reconstructed'  Spectral Energy  Distribution (SED) of \grs~at
low  energies -~as expected within the framework of the \mbox{SSC} model.
The significant intergalactic absorption of sub-TeV \grs~ leads to
both the shift of the position ($E \simeq 2 \, \rm TeV$ ) and increase of the flux 
($\nu F_\nu \simeq 10^{-9} \, \rm erg\ cm^{-2}\ s^{-1}$) of the so-called Compton peak
in the \gr~spectrum. 
An analytical approach has been recently 
proposed by (TMG) for derivation of constraints on the  jet parameters of TeV blazars.
With these revised spectral parameters, new results in  a well defined  self-consistent \mbox{SSC}
parameter-space for the jet of Mkn~501 in the high state
are obtained.

\section{Intergalactic absorption of VHE gamma rays} 

If we ignore the appearance of second-generation \gr~photons
in the source direction\footnote{This is a good approximation 
unless the intergalactic magnetic field \mbox{$B \ll 10^{-16}$~G}.
In the case of such extremely  small intergalactic magnetic fields, the new
generation  (cascade) $\gamma$-rays, produced by secondary
electrons and positrons via inverse Compton scattering on 2.7 K microwave background,
remain in the field of view of the detector and arrive almost
simultaneously with   the primary $\gamma$-rays. If so, the cascade
spectrum could dominate over the primary $\gamma$-ray spectrum,
and therefore instead   of the absorption features in the 
primary spectrum, we should detect  a
standard power-law ($\propto E^{-1.5}) $ cascade spectrum with
exponential cutoff at $E^\ast$,  determined from the condition 
$\tau(E^{\ast})=1$. },
the extinction of gamma radiation is reduced to a simple  absorption 
effect, described by an energy dependent optical
depth: $$J_{ obs}(E)=J_0(E) \, \exp{[-\tau_{\gamma \gamma}(E)]}$$
 where $J_0$ and $J_{ obs}$ are the intrinsic (source) and detected \gr~fluxes, respectively. 
In order to calculate the optical depth for $\gamma$-rays from several hundred~GeV to
20~TeV (the typical energy range of current Cherenkov telescopes), 
one needs to know the distance to the source 
$d=c \, z /H_0$, where $H_0$ is the Hubble constant, 
and the flux of the diffuse background 
in a rather broad interval from  sub-micron to 100~$\mu$m wavelengths
(for close sources, $z \ll 1$, the CIB flux is independent of $z$).
Note that if, for a given redshift  $z$, the uncertainty in the estimate  
of the  distance to the source does not exceed a factor of 1.5 
(the current estimate of the  Hubble constant $H_0$ is between 50 and 75 km/s/Mpc), 
the uncertainty connected with  the CIB flux
(directly measured and/or predicted by different cosmological models)  
is significantly  larger. However, since in the method described below  
we intend to obtain {\it independently}  the density of the CIB using  
the \gr~data,  we need to know the spectral shape (or possible shapes) of the CIB,
rather than the absolute value of the CIB,
in a broad band of  wavelengths from 0.3~\mic~to 100~\mic.

\subsection{CIB models}

The calculations of CIB  present a serious theoretical challenge
because they require a complex  treatment of number 
of cosmological assumptions and key physical processes of 
galaxy formation (see~{\it e.g.} MacMinn \& Primack 1996; 
Salamon \& Stecker 1998, Madau 1999, Primack~et~al. 1999).
It is  important  for our further discussion 
that most of  cosmological models give rather similar shapes of 
the  basic,  the `stellar' and `dust' components of radiation
- two distinct bumps at 1-2~\mic~and 100-200~\mic~(see~{\it e.g.} 
Dwek~et~al. 1998; Primack~et~al. 1999). 
The shape (the precise position, depth and width) 
of the mid-infrared `valley'  contains more uncertainties because
of the   lack of adequate information  necessary 
for  modeling the radiation  associated   with warm dust  component. 

In this paper, we use a sample of recent models providing
information from roughly 0.1 to 300~$\mu$m.
The predictions of two such models, LCDM (Lambda Cold Dark Matter)
and  CHDM (Cold and Hot Dark Matter)
suggested by Primack~et~al. (1999)  for the density of CIB is shown in Fig.\ref{fig_cirb_pct}
(top panel). For comparison we show also the maximum and minimum 
CIB fluxes derived  phenomenologically by  Malkan \& Stecker (1998)
for wavelengths $\lambda \geq 3.5 \mu \rm m$. 
However, in this paper we are primarily interested 
in absorption of low energy, $E \leq 2 \, \rm TeV$ 
$\gamma$-rays, which effectively interact with 
NIR background photons. Therefore  
below we will not use the CIB models of  Malkan \& Stecker (1998). 

In the bottom panel we present another three CIB models
suggested by  Dwek~et~al.~(1998) which differ from each other 
by cosmic star formation histories. The   
PFI and  PFC models are based on  different 
chemical evolution scenarios
directly related to the star formation rate based on 
the Pei \& Fall~(1995) calculations. The ED  model (Extragalactic DIRBE results)
is derived from UV and optical observations and reproduce the COBE fluxes
(for details see Dwek~et~al. 1998).

The flux dispersion between the curves in Fig.\ref{fig_cirb_pct} shows that neither 
the shape nor the  intensity  can be  precisely predicted 
by current CIB models.  Actually in our approach we intend to 
``measure'' the CIB density ourselves. Therefore we are interested 
in the shape rather than in the absolute flux  predicted by CIB models. 
Fortunately, in the most informative  parts of the gamma-ray 
spectrum, namely below 1~TeV and above 10~TeV, the absorption of
$\gamma$-rays from relatively nearby BL~Lac object Mkn~501 is 
contributed  mainly by two well separated parts of the 
CIB  spectrum  - $\lambda \leq $ few \mic ~and $\lambda \geq $ 10 \mic, respectively.
This allows us to introduce a 
 scaling factor $SF$, which does not change the shape,
but varies the  absolute flux of CIB within reasonable limits,
and thus to derive  the ``best-fit'' values or upper limits on
$SF$ for each specific CIB model.

\subsection{$\gamma-\gamma$ pair production}

\begin{figure}
\includegraphics[width=9cm]{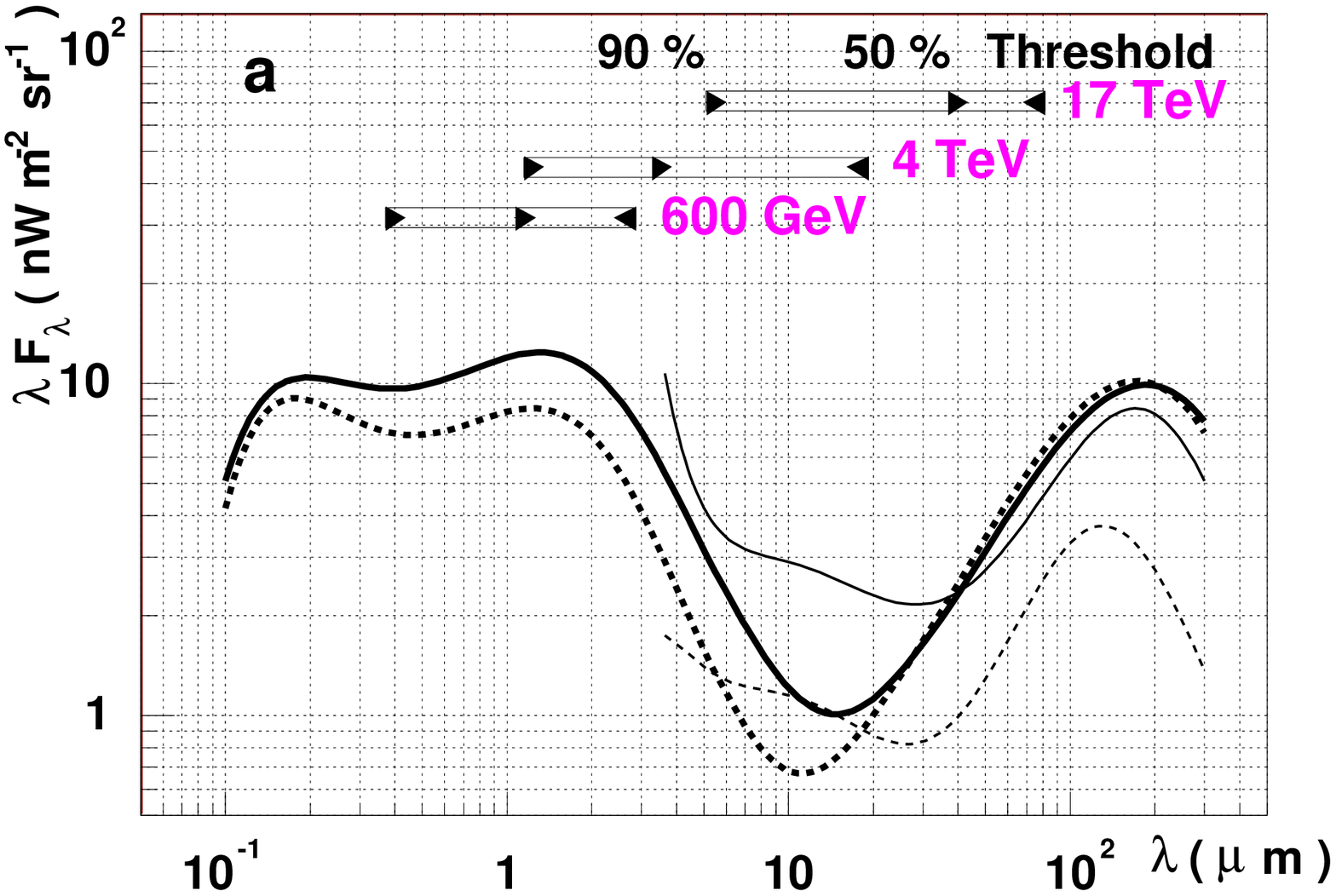}
\includegraphics[width=9cm]{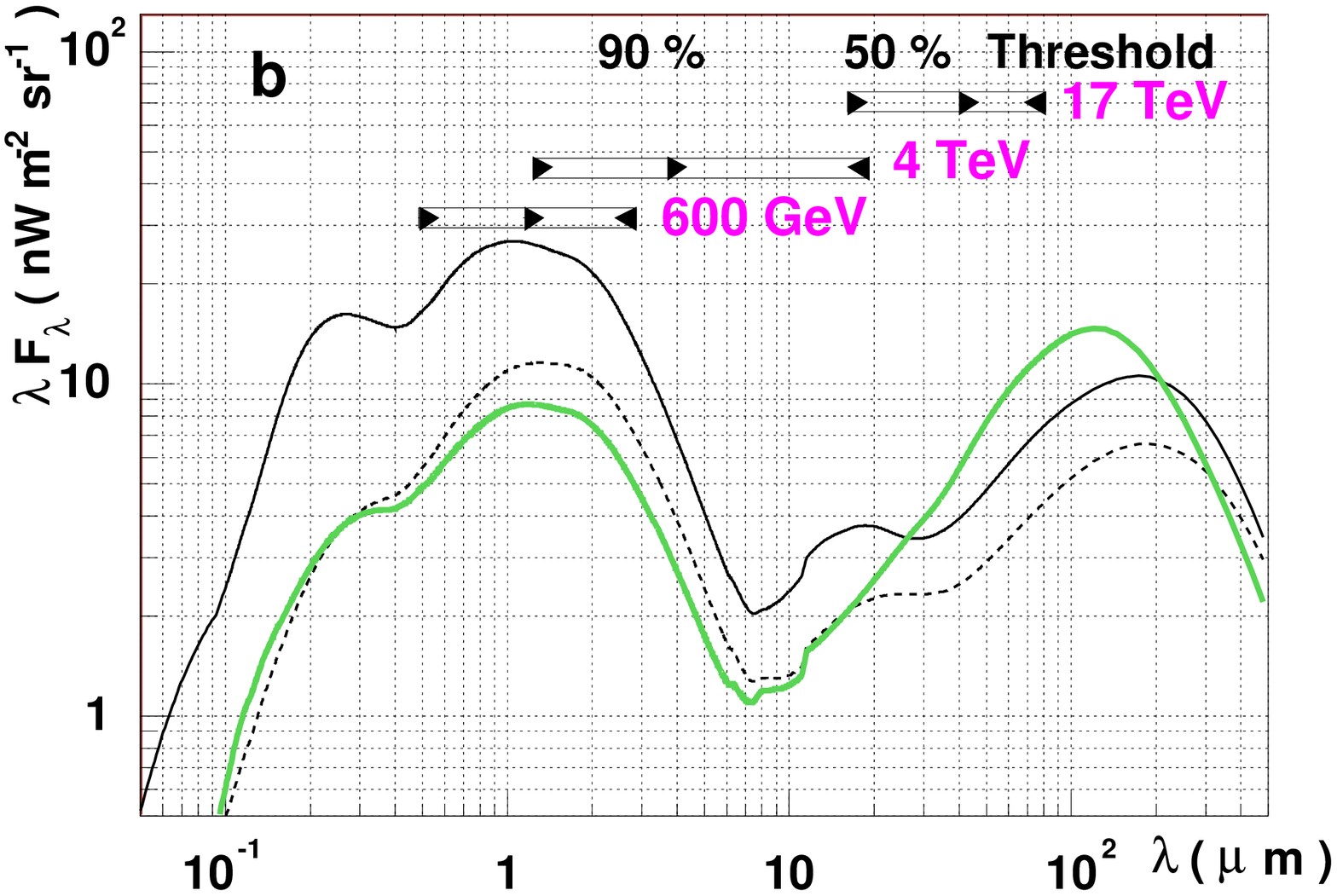}
\caption{Predictions for the CIB fluxes by different cosmological models. 
{\bf a}:
the LCDM and CHDM models of Primack~et~al. (1999) 
are shown by  thick solid and dashed lines, respectively.  
The maximum and minimum CIB fluxes
from the model of  Malkan and Stecker (1998) are shown by   
thin solid and dashed lines, respectively.
The arrows  show the wavelength intervals (from the threshold) of CIB photons
with  $50 \%$ and $90 \%$ contributions to the  
absorption of TeV  $\gamma$-rays with energies 600~GeV, 4~TeV, and 17~TeV 
computed for the LCDM distribution.
{\bf b}: three models discussed by  Dwek~et~al. (1998) 
are drawn with solid line
(PFI model),  dashed (PFC model) and grey line (ED model).
The arrows  show the wavelength intervals  computed for the ED distribution.
The thresholds  indicated by the left side of the arrows is independent of
the CIB model.
}\label{fig_cirb_pct}
\end{figure}

A \gr~photon with energy $E$ penetrating through an isotropic field of photons,
can interact with any background photon of energy 
$\epsilon \geq  \epsilon_{ th} = (m_{ e} c^2)^2/E \simeq 0.26 \,  (E/1 \, \rm TeV)^{-1} \, \rm eV$. 
Since the cross-section of the pair production   
$\gamma \gamma \rightarrow e^+e^-$ peaks at 
$\epsilon_{ max} \simeq 4   \epsilon_{ th}$ with 
$\sigma_{\gamma \gamma} \simeq 10^{-25} \, \rm cm^2$
(see {\it e.g.} Herterich 1974),  for a large class of relatively 
flat and smooth  ({\it e.g.} power-law) spectra  
of the field photons,   approximately half of 
the optical depth $\tau_{\gamma \gamma}$ 
is contributed by a rather narrow band of the background
radiation  within $\epsilon_{ max} \pm 1/2 \epsilon_{ max}$.
For a more realistic  shape of the CIB photons with two 
distinct bumps at NIR and FIR, the relative contributions of 
different parts of the spectrum of CIB do not follow this simple   
relation. In Fig.~\ref{fig_cirb_pct}, we show the energy intervals from the threshold
$\epsilon_{ th}$ of  CIB photons contributing
to the $50 \%$ and $90 \%$ of the optical depth of intergalactic
absorption calculated for the LCDM and the ED type CIB spectra 
for 3 energies of primary $\gamma$-rays:  $E=$600~GeV, 4~TeV, and
17~TeV. It is seen that $50 \%$ of absorption of 600~GeV \grs~is 
caused by a narrow band of CIB between 1-3~\mic, while the
remaining contribution comes essentially from the 0.4-1~\mic~band for the LCDM model.
The thresholds depends only on the TeV photon energy, so they are independent of the CIB model.
The bands are significantly narrower for the ED model because both NIR and FIR parts are more ``peaked''.

The absorption of multi-TeV \grs~is equally contributed both from 
mid- and far-IR parts of the spectrum. So conclusions based on the absorption
of these energetic \grs~ depend strongly on the model of the CIB, first of all
on the depth of the mid-IR `valley'  of the CIB.  
It is important to note that conclusions concerning the
change of the spectral shape of sub-TeV \grs~due to the CIB 
absorption is independent of the density of mid- and far-IR photons,
 these photons being beyond the $\gamma-\gamma$ interaction threshold.

The optical depth is a function of the $\gamma$ photon energy $E$ defined as, for a source at a redshift
$z_s \ll 1$ :
$$
\tau(E) =  \frac{c z_s}{H_0} \int_{-1}^1 \frac{1-cos \theta}{2}  d(cos \theta)  \int_{\epsilon_t}^\infty
 n(\epsilon) \sigma(E,\epsilon,\theta) d\epsilon 
$$
where $\theta$ is the angle between both photons, $\epsilon$ the IR photon energy
 and $\sigma(E,\epsilon,\theta) $ is the
pair-production cross-section.

\begin{figure}
\includegraphics[width=9cm]{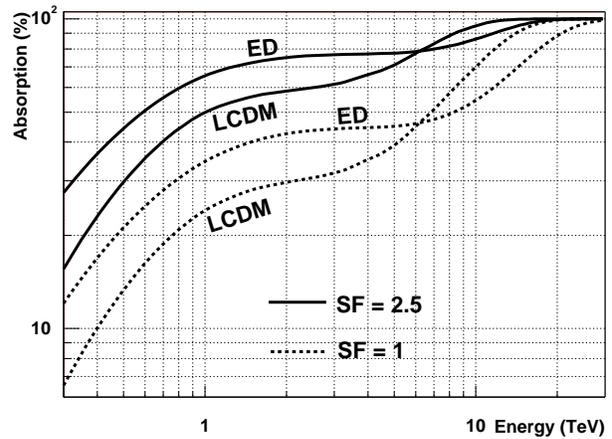}
\caption{Absorption in percents of \grs~from Mkn~501 in function 
of the gamma energy calculated 
for two different scaling factors (1 and 2.5) and two models (LCDM and ED).
}\label{fig_abs}
\end{figure}

In Fig.~\ref{fig_abs} we present  the portion (percentage) of absorbed $\gamma$-rays, 
$P=1-\exp[-\tau_{\gamma \gamma}(E)]$,  
from Mkn~501 for the LCDM and ED models assuming  
two different   scaling factors: $SF=1$ and 2.5.
Hereafter all calculations are performed
for the Hubble constant $H_0=60 \, \rm km/s \, Mpc$.

\section{Constraints on CIB} 

Given the lack of reliable information
on the intrinsic \gr~ spectrum, an upper limit on the CIB flux could be derived 
by formulation of {\it a priori}, but astrophysically meaningful requirement on the
shape of the spectrum of \grs~produced in the source.  In particular, we may require 
that  within any reasonable model of $\gamma$-ray production, and for a given
observed  \gr~spectrum  $J_{ obs}(E)$,
the {\em source}  spectrum, 
$J_0(E)=J_{ obs}(E) \, \rm exp[\tau_{\gamma \gamma}(E)] $, 
should  not contain, at any energy $E$,  a strongly ({\it e.g.} exponentially) 
rising feature. 

\begin{figure}
\includegraphics[width=9cm]{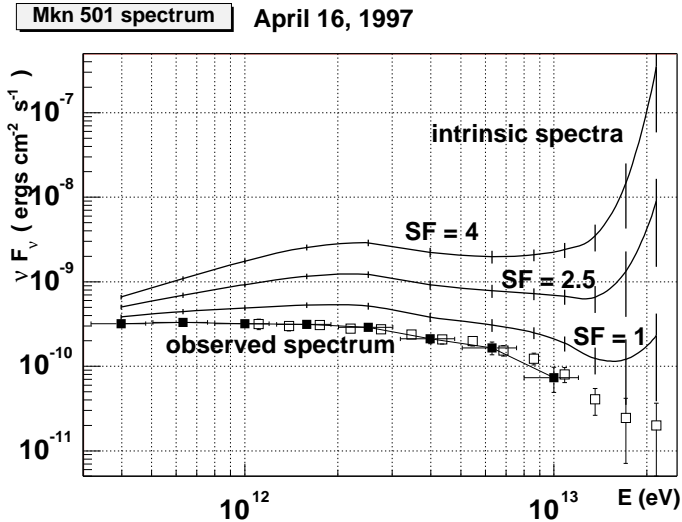}
\caption{Spectral energy distribution of \grs~from Mkn~501 as detected by  CAT 
during the April~16,~1997 flare (filled squares).  The spectral points of the HEGRA 
`time-averaged' spectrum normalized to the CAT flux at 1~TeV 
are also presented (open squares). Statistical errors only are shown.
Three intrinsic \gr~spectra are computed using the 
LCDM model and scaling factors of 1, 2.5 and 4.
}\label{fig_fir}
\end{figure}

In Fig.~\ref{fig_fir} we show the spectrum of Mkn~501 as measured by the CAT telescope
during the strong April~16,~1997 flare in the energy region from 300~GeV to 12~TeV 
(statistical errors only).
Unfortunately, because of bad weather this remarkable flare  
could not be observed  by the HEGRA telescope system. However, the
HEGRA observations of Mkn~501 revealed 
that, despite dramatic flux variations in time-scales $\leq 1 \, \rm day$, the 
shape of the energy spectrum above 1~TeV remained essentially stable  
throughout the entire state of source high activity in 1997. So we show the 
HEGRA `time-averaged' spectrum (Aharonian ~et~al. 1999) normalized to the
CAT April~16,~1997 spectrum at $E=1 \, \rm TeV$  with the 
re-scaling   factor of $\approx$2.2.  
At TeV energies the 
agreement between the CAT and HEGRA spectra is quite impressive. 
Below 1~TeV we show only the CAT spectral points since in this energy region 
both the statistical and systematic errors of the data 
obtained close to the energy threshold of the  HEGRA 
telescope system are rather large.
Moreover,  the inclusion of the HEGRA `time-averaged' spectral points 
at sub-TeV energies in Fig.~\ref{fig_fir} cannot be justified because 
of  reported noticeable variation of the  spectrum of Mkn~501 
at such {\em low } energies in the time-scales $\leq 1 \, \rm day$, 
especially during strong flares (Djannati-Ata\"{\i} ~et~al. 1999).     

\subsection{Limit on the CIB at FIR}

More than half of the intergalactic absorption of highest energy 
photon above 10~TeV detected from Mkn~501
is contributed by 
interactions with the `dust' component of CIB at wavelengths  
$\lambda \geq 10 \mu \rm m$ (see Fig.~\ref{fig_cirb_pct}).
The optical  depth  increases with energy so rapidly
that requiring the absorption-corrected spectrum to be concave is already sufficient
to impose an interesting constraint on the CIB. Therefore the most robust 
upper limits on CIB at FIR are provided by the HEGRA data above 10~TeV.   
Indeed, in this energy region the  source spectrum `reconstruction factor' 
exp$(\tau_{\gamma \gamma})$ should not exceed the exponential 
term of the observed spectrum of Mkn~501 which in the energy region 
up to 24~TeV is well  described as 
$J_{ obs}(E) \propto E^{-1.92} \exp (-E/6.2 \, \rm TeV)$  
(Aharonian ~et~al. 1999).
The intrinsic (reconstructed) spectra of Mkn~501 are shown in 
Fig.~\ref{fig_fir} 
assuming that the spectral shape of CIB is described by the LCDM model 
with 3 different scaling factors: $SF=1, 2.5$, and 4.

In order to obtain an upper limit  on the CIB intensity 
around 50~$\mu$m,  we require that the  intrinsic flux at
17~TeV (the highest 
energy point in the HEGRA data set with adequate
statistical significance;  Aharonian ~et~al. 1999)
should not exceed the flux at 4~TeV. 
In the case of the SSC model  this implies that for 
any reasonable combination of model parameters,   
the Compton peak should not appear at energies 
beyond several TeV (in particular, due to the 
Klein-Nishina effect). For the hadronic 
(``$\pi^0$-decay'') models ({\it e.g.} Dar \& Laor 1997) 
this implies that the spectrum of accelerated protons
should not be flatter than the ``nominal'' $E^{-2}$ spectrum.

It is seen  from Fig.~\ref{fig_fir} that 
the LCDM type spectrum of CIB  with scaling factor 
up to $ SF=2.5$ do not contradict this   
`reasonable-source-spectrum' criterion.
Actually, the  acceptable  range of   $SF$ 
at $68 \%$ C.L.  extends to $ SF=3.4$ (see Tab.~\ref{tab_res_FIR}).
Note that as the above formulated 
$\nu F_\nu(17 \, \rm TeV)/\nu F_\nu(4 \, \rm TeV) \leq 1$
criterion could be treated as a tight condition, 
the results presented in  Tab.\ref{tab_res_FIR} are based on  
a rather  relaxed,  $68 \%$ C.L.  statistical requirement.

The upper limits on $SF$ 
are quite sensitive to  the exact shape  of the CIB spectrum 
because the  absorption of 17~TeV $\gamma$-rays is 
contributed mostly by the background  photons from the 
rising branch of the CIB spectrum ($\lambda \geq  10$ \mic, 
see fig.~1), therefore the optical depth increases 
strongly with energy of $\gamma$-rays.  

The results are 
summarized in Tab.~\ref{tab_res_FIR}.
The derived values of $SF_{\rm  max}$ give upper limits on 
the absolute flux of  CIB from 20 to 80~\mic~ shown in Fig.~\ref{fig_new_CIRB}. 
Note that the CHDM, LCDM, and ED models have rather similar slopes 
between 20 and 80~$\mu$m  (see Fig.~\ref{fig_cirb_pct}).   
Therefore the upper limits on CIB corresponding to these three models are 
very close. The same is true for  the PFI and PFC models. 
For that reason in Fig.~\ref{fig_new_CIRB} we show only  
upper limits on CIB  flux corresponding to LCDM 
(black line) and PFI (grey line) models as 
representatives of these two group of models, respectively.

\begin{table}[hhh]
\begin{center}
\begin{tabular}{cccc}
\hline
Model & $SF_{max}$ & Upper Limit & (nW m$^{-2}$ sr$^{-1}$) \\
 & & at 20~$\mu$m & at 80~$\mu$m\\
\hline
LCDM & 3.4 & 3.6 & 21.1 \\
\hline
CHDM & 2.9 & 2.9 & 18.0 \\
\hline
PFC & 2.6 & 5.9 & 11.4\\
\hline
PFI & 1.7 & 6.3 & 12.7\\
\hline
ED & 1.15 & 2.9 & 14.2 \\
\hline
\end{tabular}
\end{center}
\caption{
The 68~$\%$ C.L 
upper limits on the scaling factor $SF$ 
derived at  FIR from 20 to 80 \mic~ 
for different  CIB models.}
\label{tab_res_FIR} \end{table}

Finally we note that although 
the extrapolation of the FIR upper limits shown in Fig.~\ref{fig_new_CIRB}
do not contradict the COBE measurements at $\lambda \geq 100$ \mic, 
the maximum energy of about 17~TeV reported from Mkn~501
with high statistical  significance, is not sufficient to
provide model-independent information about the CIB at wavelengths 
beyond 80~\mic.

\subsection{Limit on the CIB at NIR}

At shorter wavelengths, $\lambda \leq 10~\mu$m,  the  
constraints on the CIB come mainly 
from data below 10~TeV.
For a power-law spectrum of the
CIB, $n(\epsilon) \propto \epsilon^{-\beta}$, the optical 
depth is proportional to  $\tau_{\gamma \gamma} \propto E^{\beta-1}$
(see {\it e.g.} Gould and Schreder 1967).
In the  2-10~$\mu$m side of the `valley'  the CIB,  contributed mainly
by the starlight,  has a typical spectrum
$\lambda F_\lambda \propto  \lambda^{\delta}$ with $\delta \simeq -1$
(see {\it e.g.} Dwek ~et~al. 1998;  Primack ~et~al. 1999),
or $n_{ CIB}(\epsilon)  \propto \epsilon^{-1}$. Therefore  the shape of the 
$\gamma$-ray  spectrum at energies between 2~TeV and 10~TeV
  remains  essentially unchanged  ($\tau_{\gamma \gamma}(E) \simeq$~constant),   
although the  absolute absorption  effect could be very  large.
This effect is clearly seen in Fig.~\ref{fig_abs}. It  makes rather difficult to extract information 
about CIB  based merely on the \gr~data in the~{\it intermediate} 
energy range of $\gamma$-rays  between 2 and 10~TeV. 

Essentially  more information 
about the CIB is  contained in  {\it low} energy,  
$E \leq 1-2 \, \rm TeV$   $\gamma$-rays,
since in this energy region the absorption effect 
again becomes energy-dependent
(see Fig.~\ref{fig_abs}).  For the redshift  of  Mkn~501, $z=0.034$, 
the optical depth $\tau_{\gamma \gamma}$ at 500~GeV remains less than 1 for any 
reasonable assumption about the 
CIB at NIR. At 1-2~TeV the optical depth could be close  
or even exceed 1,  if $ SF \geq 2$. 
Therefore,  a  variation of the density of the CIB
by the scaling factor $SF$ from 1 to 4  
could lead to dramatic (up to factor of 10) increase of the `reconstructed' flux
at 1-2~TeV, while at energies below 500~GeV the impact is still not  very large 
(less than factor of 2). This implies that the slope of the \grs~energy spectrum 
from Mkn~501 below 1~TeV is very sensitive to the level of the CIB flux at  NIR. 
Therefore it  can be used not only to constrain 
the CIB, but  also to ``measure'' the CIB flux   
at wavelengths from 0.25 to few \mic,  provided 
that a reliable intrinsic \gr~spectrum  could be derived from 
multi-wavelength studies of the source. Note that the result is insensitive 
to the FIR density.

The April~16,~1997 flare of Mkn~501 is particularly well suited for this task because 
(i) the appropriate distance to the source of about 170~Mpc - {\em sufficiently} large to 
provide non-negligible ({\it e.g.} measurable  absorption at  $E \leq 1$~TeV), 
and at the same time {\em not too} large for  strong suppression of TeV emission;  
(ii) the availability of high quality X-ray and sub-TeV \gr~data obtained simultaneously
by BeppoSAX and CAT, and  (iii) the general belief that the TeV radiation of this source
has a synchrotron-self-Compton origin  (\mbox{SSC}).  In the \mbox{SSC} 
scenario the TeV \grs~and synchrotron X-rays in a broad band from 0.1~keV to
$\geq 100 \, \rm keV$  are produced by the same population of electrons
accelerated in the relativistic jet with Doppler factor $\geq 10$
(for a review see Sambruna 1999).  If the \mbox{SSC} model works for this source,
it is possible to robustly predict  the intrinsic \gr~spectrum based
on the multi-wavelength observations during  a strong flare of Mkn~501.        

In this paper we do not intend to model the April~16,~1997
flare, but rather we use a few well known results  from previous 
applications of the \mbox{SSC} model. 
Within this,  for a realistic set of parameters characterizing 
the synchrotron X-ray and inverse-Compton (IC) \gr~
emitting  jet,  the \gr~photons with energy 
$\leq 1 \ TeV$ or so are produced in the 
Thomson regime, and therefore are described 
by the same spectral index as the X-rays do below the so-called break
(see {\it e.g.} Tavecchio ~et~al. 1998).  The spectrum of X-rays of  
April~16,~1997
flare measured by  BeppoSAX (Pian ~et~al. 1998)  
had a photon index 1.5 at 1~keV, and remained unusually hard 
up to very high energies with photon index 1.7 at 100~keV.
Although  for more definite conclusions one needs comprehensive modeling
of the flare in a broad frequency band,  we suppose that 
the sub-TeV \gr~spectrum should repeat the  spectral shape of 
relatively low energy X-rays, so we  expect a photon index of intrinsic 
spectrum of sub-TeV \grs~close to 1.5 (see {\it e.g.} Krawczynski ~et~al. 1999 
for more general discussion).  Another characteristic feature of the 
\gr~ spectrum is its so-called `Compton peak' which unavoidably   appears 
in any \mbox{SSC} or, more generally,  external radiation Compton
models (see {\it e.g.} Sikora~et~al. 1997).  

The spectrum of the IC  $\gamma$-rays in a broad 
energy band, including the   
transition region from the Thomson regime  to the Klein-Nishina 
regime (which is around 1~TeV here),  has a rather complicated form.  
For convenience, following the recommendation of
Sikora ~et~al. (1997), we fit 
the \gr~spectrum,  corrected for the intergalactic absorption
with a given scaling factor  $SF$, 
on the ($log(\nu F(\nu)),log(E)$) plane by a parabola
with a slope at 600~GeV $\alpha_{600}$ (photon index). 

\begin{figure}
\includegraphics[width=9cm]{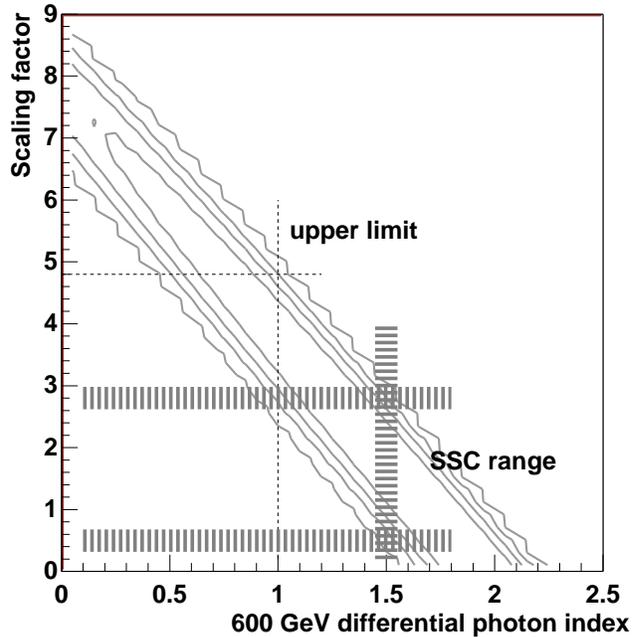}
\caption{Relation between the scaling factor 
$ SF$ and the photon index of the \gr ~spectrum
after correction for absorption in the CIB, 
$\alpha_{600}$,  
obtained from the $\chi^2$ fit 
for the LCDM model. The contour levels are at 10, 90 and 99~$\%$~C.L.; the scaling factor 
related to the upper limit is the maximum scaling factor  compatible at 99~$\%$~C.L. with 
$\alpha_{600}=1$ while the scaling factors allowed by the SSC model 
lie between the minimum and the maximum scaling factors  compatible at 99~$\%$~C.L. with 
$\alpha_{600}=1.5$.
}\label{fig_chi2}
\end{figure}

The relation between $SF$ and $\alpha_{600}$ 
computed  by $\chi^2$ minimization  is shown in Fig.~\ref{fig_chi2}. 
These results imply  that  the  photon index 
expected around 1.5  within \mbox{SSC} model for the `Thomson'
branch of the IC spectrum  could be  
achieved for a \mbox{scaling} factor 
$SF$ ranging between 0.5 and 2.8 at 99$\%$~CL (LCDM model).
This implies a  CIB flux of $\approx$5-35~nW~m$^{-2}$~sr$^{-1}$  at 1~$\mu$m. 
The analogous calculations  for each of the other four CIB models
lead to the limits summarized in Tab.~\ref{tab_res}.
Remarkably, the difference  between the results derived for 
different models  is always less than 20-30~\%. 
Thus  the conclusion about the lower and upper 
limits of variation of the CIB intensity at 1~\mic, 
allowed by the SSC model, is very robust.

The above estimate of the CIB flux is based on a model-assumption that
the observed sub-TeV \grs~are produced within the 
framework of the \mbox{SSC} model as a result of inverse Compton 
scattering of highest energy electrons responsible also for the 
synchrotron radiation of the jet. However,  
the strong relation between the scaling factor $SF$ and 
the photon index of the intrinsic sub-TeV \gr~spectrum
 allows very robust constraint on the CIB without 
a specific  model assumption. Indeed from Fig.~\ref{fig_chi2} follows that at large 
scaling factor $ SF$, the `reconstructed' \gr~spectrum becomes
extremely hard, {\it e.g.} with $\alpha_{600} \leq 1$ for $ SF \approx 5$.
Such spectrum seems to be unacceptably hard  for any 
realistic scenario of \gr~production connected with either protons or electrons. So
we may draw a conclusion that at $99 \, \%$~C.L.   
the value $\simeq 60\ \rm nW/m^2 sr$ should be considered
as an absolute  upper limit on the CIB flux at 1~\mic.   
An assumption of  a smaller distance to Mkn~501, {\it e.g.} 
 adopting  for the Hubble constant 
$H_0$=75~km.Mpc$^{-1}$.s$^{-1}$ (instead of 60~km.Mpc$^{-1}$.s$^{-1}$),
could soften  this upper limit only by $25 \, \%$. 

\begin{figure}
\includegraphics[width=9cm]{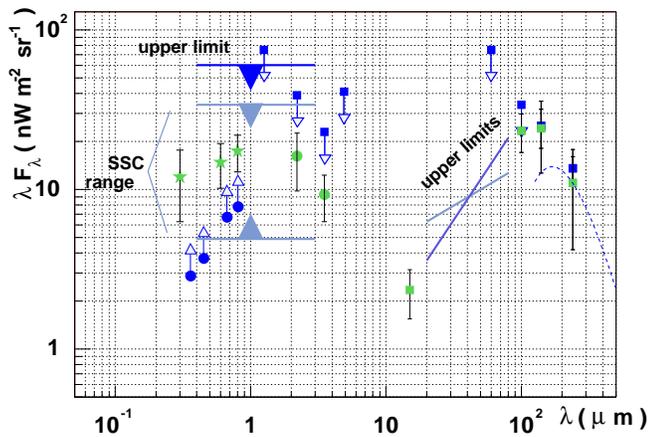}
\caption{CIB spectral energy distribution. 
The results of this paper are represented by large bars (0.4-3~$\mu$m, 99$\%$ CL) 
and are valid for all reasonable models or
the SSC model (grey bars). Two limits are plotted in the range 20-80~$\mu$m 
(68$\%$ CL) as the results depend on the CIB slope. See text for details. 
The DIRBE results  are shown by black squares (Hauser~et~al. 1998).
Three  grey  squares correspond to CIB density extracted from WHAM H$_\alpha$ survey and
 Leiden/Dwingeloo HI data  at 100,
140 and 240~$\mu$m  (Lagache~et~al. 1999). The grey square
at 15~$\mu$m comes from galaxies counting (Elbaz et al. 1998).
 The two grey dots show the fluxes of 2.2 and 3.5~$\mu$m
 (Dwek \& Arendt 1998, Gorjian~et~al. 1999).
The lower limits correspond to the Hubble Deep Field  galaxy counts (Pozzetti~et~al. 1998),
and the grey stars come from HDF results combined with
ground-based spectrometry (Bernstein~et~al. 1999).
The range of the CIB flux detected by FIRAS at very long wavelengths is shown by  
dotted line (Fixsen~et~al. 1998).
}
\label{fig_new_CIRB}
\end{figure}

In  Fig.~\ref{fig_new_CIRB}  we  compare the  limits on the CIB intensity 
derived above  with the direct measurements.
Results obtained in this paper  are shown by large horizontal bars
corresponding to the energy range of
CIB photons which can interact with 600~GeV photons.
The 99$\%$~CL limits correspond to  the 5  different CIB models.
The upper limit at 60~nW~m$^{-2}$~sr$^{-1}$ assumes 
a differential source spectrum
softer than $E^{-1}$. The allowed range 
of variation of the CIB between 5 and 35 ~nW~m$^{-2}$~sr$^{-1}$ 
is derived assuming inverse Compton origin of   TeV radiation of Mkn~501.   
These results are in good agreement with  recently reported 
fluxes of CIB at 2.2~$\mu$m and 3.5~$\mu$m  based on the 
DIRBE measurements (Dwek \& Arendt 1998, Gorjian ~et~al. 1999),
as well with the measurement of the background radiation
from absolute photometry  (Bernstein~et~al. 1999).

\begin{table}[hhh]
\begin{center}
\begin{tabular}{ccc}
\hline
Model & Upper Limit & SSC limits \\
 & nW m$^{-2}$ sr$^{-1}$ & nW m$^{-2}$ sr$^{-1}$ \\
\hline
LCDM & 48 & 5 to 28 \\
\hline
CHDM & 60 & 7 to 34 \\
\hline
ED & 51 & 7 to 30\\
\hline
PFC & 44 & 5 to 26 \\
\hline
PFI & 57 & 6 to 34 \\
\hline
\end{tabular}
\end{center}
\caption{The upper limit on the CIB based on the assumption  
that the differential  source spectrum of $\gamma$-rays 
at sub-TeV energies is softer than $E^{-1}$, and the interval 
of CIB densities allowed by the SSC model of TeV radiation of 
Mkn~501. The 99~$\%$ C.L. limits are relevant  to the  NIR wavelengths   
from 0.4  to 3~$\mu$m, 
corresponding to the energy range of interaction for 600~GeV photons.}
\label{tab_res} \end{table}

\section{Impact of the CIB flux on the  jet parameter-space of Mkn~501} 

\begin{figure}
  \includegraphics[width=8cm]{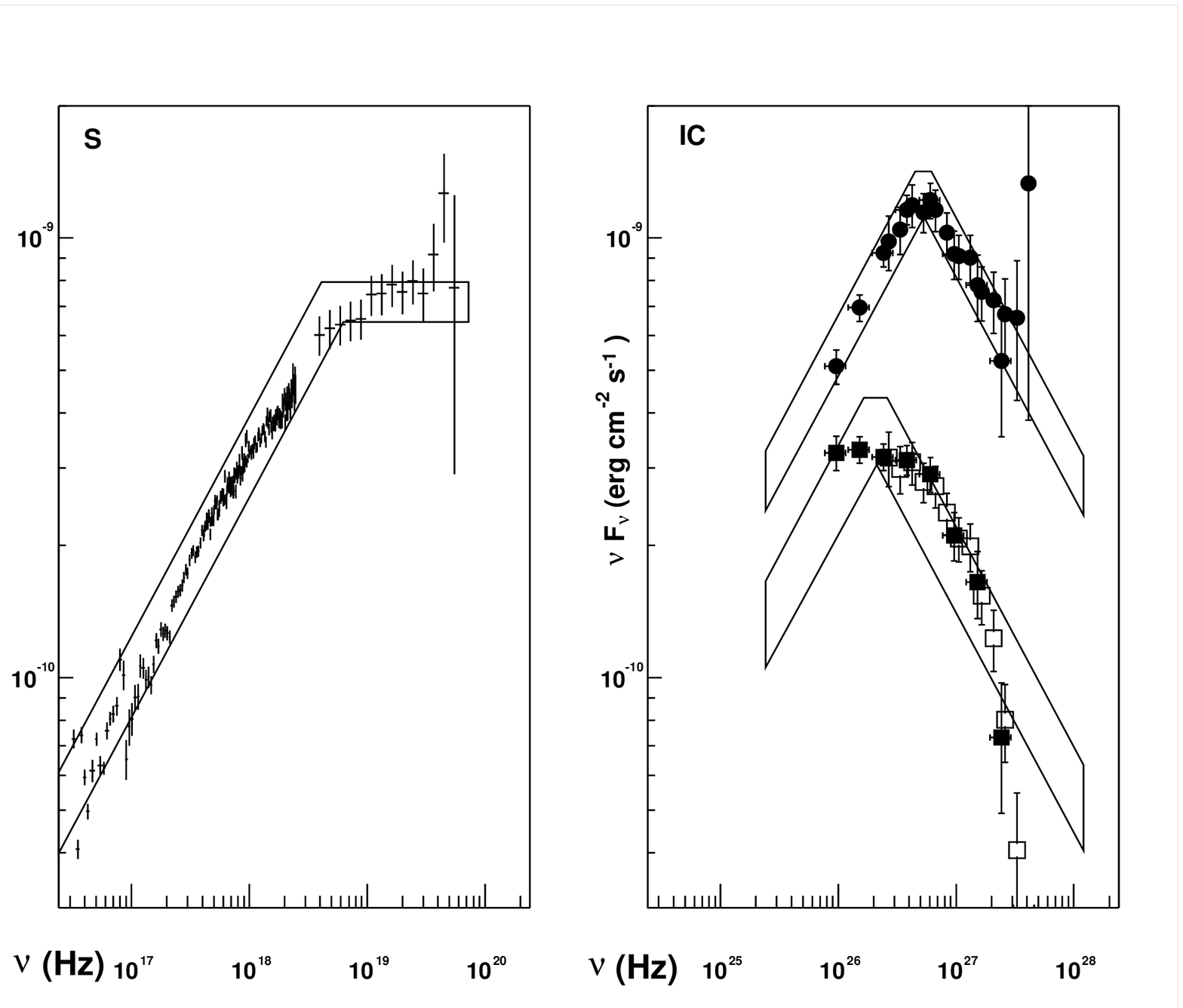}
  \caption{Spectral energy distribution of the April~16,~1997 flare of Mkn~501. 
The X-ray data are from BeppoSAX observations (Pian ~et~al. 1998).
The  \gr~fluxes measured by CAT are shown by filled squares.
The HEGRA `time-averaged' spectral points obtained during
the entire  1997 outburst of Mkn~501 (Aharonian ~et~al. 1999) 
are also shown (open squares). Statistical errors only are drawn. 
The filled dots correspond to the 
 CIB absorption-corrected fluxes  computed for LCDM model ($SF=2.5$). 
The 3~$\sigma$ are used to model the spectra. 
}
  \label{fig_sed}
\end{figure}

The analysis of the spectral shape and variability of  
the synchrotron and IC components of non-thermal radiation of 
TeV blazars within the framework of a single-zone 
\mbox{SSC} model  may yield  important  constraints 
in the parameter-space of  the X-ray and \gr~production region
(see {\it e.g.} TMG, Bednarek \& Protheroe 1999).
Among the key observables used in derivation of constraints 
on the jet parameters are the frequency $\nu_{IC}$
and the {\em apparent} luminosity  \footnote{Below for convenience we  
use the {\em apparent} luminosity of the source 
which is defined as the luminosity of an isotropically 
emitting source at a given distance.  The intrinsic luminosity of the 
blob relativistically moving towards the observer with Doppler 
factor $\delta \gg 1$ is much ($\propto \delta^4$) smaller.}  
$\nu_C L(\nu_{IC})$ of the Compton peak.
Since the intergalactic absorption of \grs~
could significantly deform the original information
about both  parameters (see Fig.~\ref{fig_sed}), we study the 
impact of the CIB density in the 
parameter-space of the jet in Mkn~501. 
For calculations, we  use the convenient analytical approach
recently developed by (TMG) for the
homogeneous \mbox{SSC} model of TeV blazars. Detailed discussion of the SSC model
is not the aim of this paper but rather we show the impact of the CIB absorption
on the derivation of these parameters.

\subsection{Model parameters and observables}
The homogeneous \mbox{SSC}  model describes the emission of electrons 
in a single blob with three basic model parameters: the radius $R$, 
magnetic field $B$, and Doppler factor of the bulk motion $\delta$ of the blob. 
An important information about the ratio $R/\delta$ is contained in the  
observed source variability time-scale $t_{ var}$, namely 
the radius of the source  should not exceed 
$R_{ max} = t_{ var} \times c \times \delta$.  Thus the 
constraints in the parameter-space for the given maximum size 
(defined through the variability time-scale)  can be expressed  
in terms of two parameters  
on the  ($\log B,\log \delta$) plane. 

A distinct  feature of the \mbox{SSC} model is the characteristic 
SED of radiation with two pronounced, synchrotron and Compton  bumps which,
in the case of X-ray selected BL Lac population, 
appear  in the X-ray and the TeV $\gamma$-ray bands respectively
(see {\it e.g.} Ulrich~et~al. 1997).
The observations of X-rays require a population of relativistic electrons
with  power-law spectrum broken at energy $\gamma_b$. The specific values of the indices
are determined by the spectral shape of the synchrotron radiation below and above the synchrotron peak.

In the quiescent state, the synchrotron radiation of Mkn~501 
is characterized by a synchrotron peak at $\nu_s \simeq 10^{16}$~Hz,
and spectral indices  $\alpha_1=0.5$ and $\alpha_2=1.75$ (TMG).
 During the April~16,~1997 flare 
the X-ray spectrum was exceptionally hard with photon  index less than 2
at least up to 100~keV, indicating a dramatic shift of the synchrotron peak
by at least two orders of magnitude (Pian~et~al. 1998). 
Due to the lack of statistics,
both the exact position for the synchrotron peak position  and 
the synchrotron luminosity during the April~16, flare are not well defined. 

The synchrotron spectrum  of Mkn~501 in the high state 
is represented in the form of broken power-law 
with  spectral indices 
$\alpha_{1}=0.5$ and $\alpha_{2}=1$
below and above the energy 
$h\nu_b=21.5 \, \rm keV$  
($\nu_b =5.2 \pm 0.3 \times 10^{18} \, \rm Hz$
(see Fig.\ref{fig_sed}).  Additionally, we assume a high energy cutoff
at 300~keV, which  naturally could be attributed 
to the cutoff in the 
acceleration spectrum of electrons. 
The index $\alpha_{2}$ as well as the position 
of the cutoff are rather qualitative,  but fortunately the final   
conclusions do not depend strongly of their exact values.
The total luminosity is then 
$\nu_s L(\nu_s)$=2.5$\pm$0.1~10$^{45}$~erg.s$^{-1}$.

According to the TMG description, the IC peak should be represented by a
broken power-law with the same spectral index $\alpha_1 = 0.5$ before the maximum and
the spectral index $\alpha_2 = 1.5$ after, as shown in Fig.~\ref{fig_sed}. 
The position of 
the maximum, and therefore the total energy depend on the assumption about the CIB absorption.
Tab.~\ref{tab_nuLnu} presents these observable parameters computed
 with no absorption and with the LCDM and PFI models, using $SF$=1 or 2.5.

It is interesting to note that the correction of the observed \gr~spectrum 
for the intergalactic
absorption,  assuming typically a CIB flux at the level of 25~nW~$^{-2}$~sr$^{-1}$ at
\mbox{$\lambda \sim 1$~\mic},  results in  a symmetric SED with a distinct
maximum  at 2~TeV, and power-law type behavior 
with spectral indices  0.5 and 1.5 below and above the peak, respectively, although
we constrained {\it only} $\alpha_1=0.5$.
This is a feature predicted  by the  \mbox{SSC} model 
for the IC radiation provided that the synchrotron X-radiation is described by a 
broken power-law with spectral indices $\alpha_1=0.5$ and $\alpha_2=1$
(TMG, Krawczynski~et~al. 1999).

\begin{table}[hhh]
\begin{center}
\begin{tabular}{lrrrr}
\hline
Model & $\nu_{IC}$~~ &$\nu_{IC}F(\nu_{IC})$ &
 $E_{IC max}$~ & $\nu_{IC} L(\nu_{IC})$\\
\hline
no absorpt. &  2.1$\pm$0.2 & 3.9$\pm$0.1 & 0.9$\pm$0.1 & 1.3$\pm$0.1 \\ 
\hline
LCDM $\times1$ & 3.6$\pm$0.2 & 6.0$\pm$0.2 & 1.5$\pm$0.1 & 2.1$\pm$0.1 \\
\hline
LCDM $\times2.5$ & 5.3$\pm$0.3 & 13.1$\pm$0.3 & 2.2$\pm$0.1 & 4.5$\pm$0.1 \\ 
\hline
PFI $\times1$ & 4.5$\pm$0.2 & 9.4$\pm$0.2 & 1.9$\pm$0.1 & 3.2$\pm$0.1 \\
\hline
PFI $\times2.5$ & 7.7$\pm$0.5 & 40.0$\pm$0.1 & 3.2$\pm$0.2 & 13.9$\pm$0.4 \\
\hline
\hline
\end{tabular}
\end{center}
\caption{The parameter 
$\nu_{IC}$ (in units $10^{26}$~Hz), 
$\nu_{IC}F(\nu_{IC})$ 
(in units $10^{-9} \, \rm erg \ cm^{-2}  \ s^{-1}$), 
$E_{IC max}$
(in units $10^{12}$~eV) and 
$\nu_{IC} L(\nu_{IC})$
(in units $10^{45} \ \rm erg \ s^{-1}$) for two CIB models
 (LCDM and PFI) and two values for the scaling factor 
$SF$ (1 and 2.5). The first line of the table corresponds to the 
$\gamma$-ray  spectrum without intergalactic absorption. }
\label{tab_nuLnu} \end{table}

\subsection{The constraint regions on the ($\log B,\log \delta$) plane}

TMG  have suggested three independent constraints on allowed 
regions in the ($\log B,\log \delta$) plane  based on (i) the 
positions of the synchrotron and Compton peak frequencies (area  A), 
(ii) the synchrotron and Compton   peak luminosities (area B),  
and (iii)  the equilibrium between radiative cooling and escape of 
electrons (area C). It is important to note that the convenient  
approximate analytical  formulae  derived by TMG are applicable
to the treatment of the IC scattering relativistic (Klein-Nishina) regime required here.

The first  condition leads to the following simple relation between $B$ and $\delta$ (eq.~16 in TMG)
\footnote{Since we do not have here 
$ \gamma_{max}$(=300\ keV) $\gg \gamma_b $(=100\ keV),
the expression of $g(\alpha_1,\alpha_2)$ must be modified.
$g(\alpha_1,\alpha_2) = exp(1/(\alpha_1-1)+1/(2*(\alpha_2-\alpha_1))*(1-\sqrt{\gamma_b/\gamma_{max}}) $ }:
\begin{equation}
(A) \ B \, \delta_{10}^{-1} \approx 1.5  \frac{\nu_{s,19}}{\nu_{IC,26}^2}
\end{equation}
where 
$B$ is in Gauss,
$\delta_{10}=\delta/10$,
$\nu_{s,19}=\nu_s/10^{19} \, \rm Hz$,  and
$\nu_{IC,26}=\nu_{IC}/10^{26} \, \rm Hz$.

\begin{figure}
\includegraphics[width=9cm]{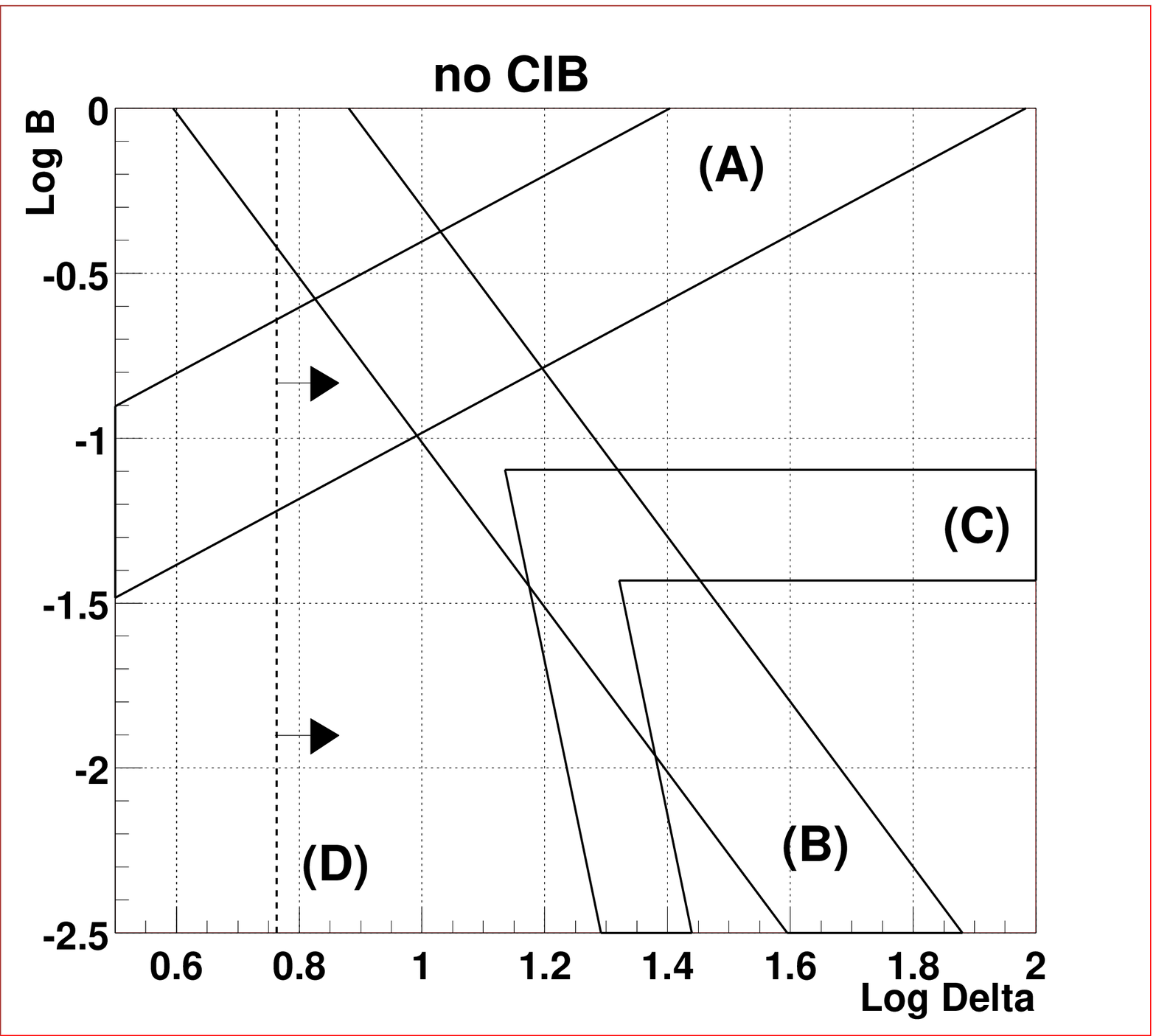}
\includegraphics[width=9cm]{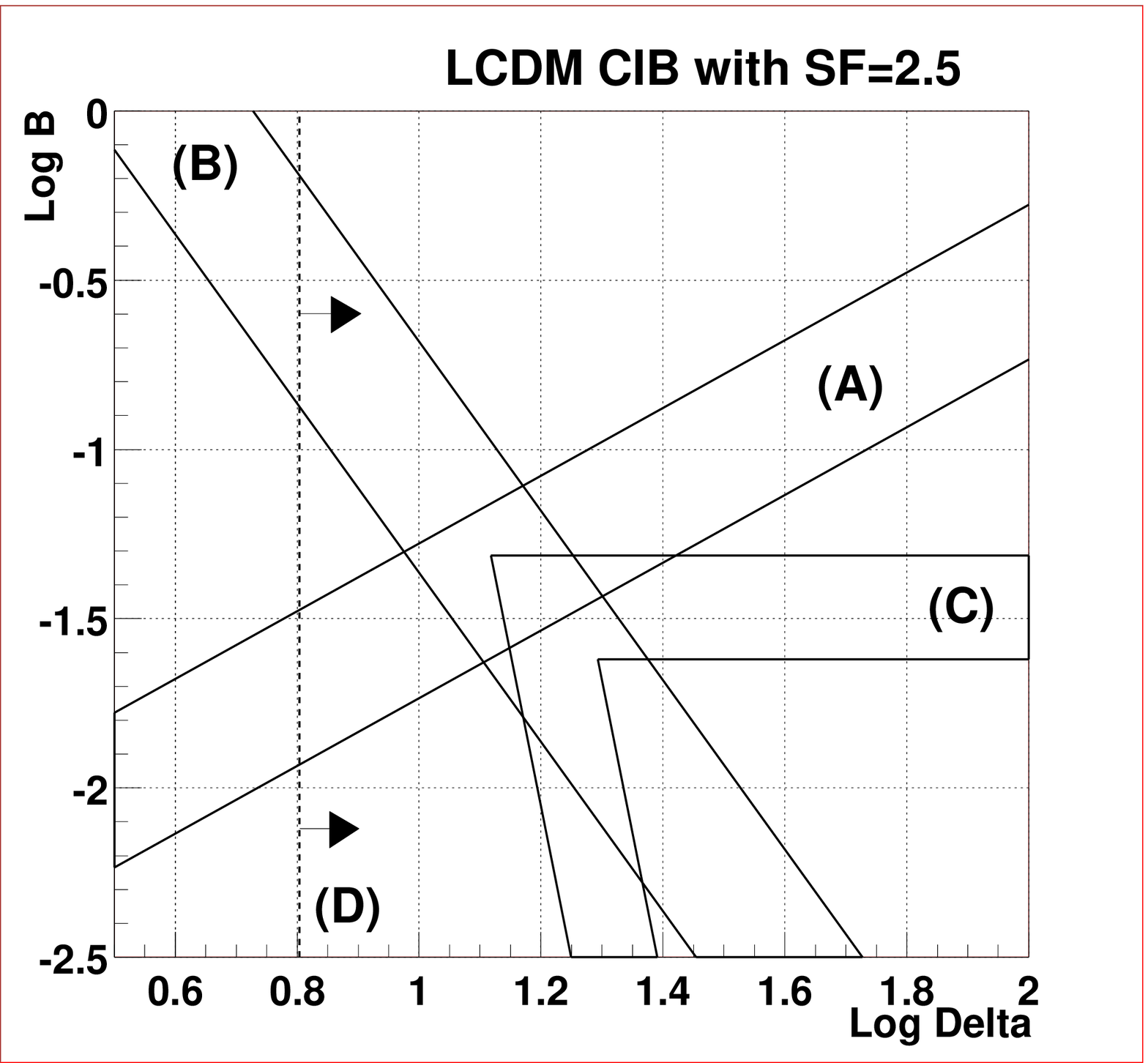}
\caption{Parameter-space for the April~16,~1997 flare of Mkn~501 (with
$t_{var}$=10h) and $SF$= 0 ({\it i.e.} no intergalactic
absorption) or $SF=2.5$  for the LCDM model. See text for details on areas (A), (B), (C) and (D).}
\label{fig_Bdelta}
\end{figure}

The IC peak is very sensitive
to the CIB absorption.
Typically, the peak position shift by a factor 3 (up to 10)
towards higher frequencies for the absorption-corrected spectrum. The total energy follow 
a similar evolution.
 This implies that 
the ignorance of the intergalactic \gr~absorption would overestimate
the  $B/\delta$ ratio by a factor of up to 10 or even more.
This is seen in Fig.~\ref{fig_Bdelta} from comparison of the regions (A) on the 
top and bottom panels. 

The comparison of the observed  luminosities  in the  
synchrotron and the IC peaks  gives the second relation between the 
magnetic field and the Doppler factor (eq.~22 in TMG)
\begin{equation}
(B) \  B  \delta_{10}^{2.5} \geq 
0.5 \, (\nu_{s,19} \ \nu_{IC,26})^{-1/4} 
\, \frac{(\nu_s L(\nu_s))_{45}}{(\nu_{IC} L(\nu_{IC}))_{45}^{1/2}} \, t_{var,h}^{-1} 
\end{equation}
where $(\nu_s L(\nu_s))_{45}=(\nu_s L(\nu_s))/10^{45} \, \rm erg\ s^{-1}$,
$(\nu_{IC} L(\nu_{IC}))_{45}=(\nu_{IC} L(\nu_{IC}))/10^{45} \, \rm erg\ s^{-1}$
are the synchrotron and IC peak {\em apparent} luminosities
 and  $t_{var,h}=t_{var}/1 \, \rm h$ is the source variability time-scale. Here $t_{var,h}$=10 is assumed.

The region (B) shown in Fig.~\ref{fig_Bdelta}  is based on Eq.~(2) with allowed 
ranges of uncertainties (3~$\sigma$) in  the positions and total energies
of the synchrotron and 
Compton peaks as described above.  
Also, it is interesting to note  that the ignorance of 
the CIB absorption of \grs~would lead to a 
conclusion that the radiative cooling of electrons 
in the jet is  well dominated by synchrotron losses. 
This,  however, could not be true, since after the correction of \gr~fluxes
for the intergalactic absorption, the IC luminosity in fact could be as high as 
(or even exceed) the synchrotron luminosity (see Fig.~\ref{fig_sed}).

A relation between $B$ and $\delta$ arises 
if one assumes that the break in the electron spectrum at $\gamma_{ b}$ is
determined from the equilibrium between the cooling and escape from the source 
(obviously, there could be other reasons for the break in the electron spectrum, {\it e.g.}
connected with the character of the acceleration mechanism). This 
model assumption leads to the following constraints in the
Compton-cooling dominated regime (eq.~32 and 34 in TMG)\footnote{
There is a mis-print in the TMG paper. The exponent of $\delta$ in Eq.~34
should be ($6/(1-\alpha_1)-3$) instead of ($6/(1-\alpha_1)$).
}:
 \begin{equation}
(C1) \  B \delta_{10}^9 \geq 37.
(\nu_s L(\nu_s)_{45})^2 \, t_{ var,h}^{-2}  \, 
\beta_{esc}^{-2} \nu_{IC,26}^{-1}
\end{equation}
and synchrotron-cooling dominated regime: 
\begin{equation}
(C2) \  B \geq 0.18 \ \beta_{esc}^{1/2}(t_{var,h} \,  \nu_{IC,26})^{-1/2}
\end{equation}

Here  $\beta_{esc}$ is the electron escape velocity 
in units of the speed of light, {\it i.e.} a parameter which describes the energy-independent
escape time as $t_{esc}=\beta_{esc} \, R/c$. 
Following TMG, we allow a change of this rather uncertain 
parameter within limits from 1/3 to 1.

To avoid strong absorption of TeV photons inside the blob due to
pair-production of $\gamma$-rays interacting with optical photons, a minimum value for $\delta$ can be
 computed ; it is used to check the validity of the area defined by the intersection between (A), (B) and (C) regions.
For the Compton peak luminosity, we obtain

\begin{equation}
(D)\ \delta > 6.6\left( \frac{ L(\nu_s)_{26}}{t_{var,h}}  
\sqrt{\nu_{IC,26} \nu_{s,19} }\right)^{0.2}
 \end{equation}

The impact of the intergalactic absorption of \grs~on the constraints 
described by the regions (B), (C) and (D)  is less dramatic than on the 
region (A). Nevertheless, as can be seen from Fig.~\ref{fig_Bdelta},  even for these regions 
the effect is not negligible,  and  should be taken into account in any realistic 
attempt to constrain  self-consistently the parameter-space 
in future studies based on detailed modeling of temporal and spectral characteristics
of radiation. The scaling factor $ SF$
which determines effectively the  flux of CIB at NIR, and therefore defines the level
of distortion  of $\gamma$-radiation in the sensitive 
region of the Compton peak, should be considered as a new free parameter in such studies.
  
Although the detailed modeling of radiation of Mkn~501 
is beyond the framework of 
this paper,  our study  shows that the inclusion of the intergalactic
absorption in the treatment results in a more consistent picture.   
In particular it moves the region (A) into the regions (B) and (C),
and thus overcomes the incompatibility of different constraints on 
the principal model parameters.
It is seen from Fig.~\ref{fig_Bdelta} that after the correction of 
$\gamma$-ray fluxes for the intergalactic absorption, there
is a trend for appearance of a compact {\em common} area for all three (A), (B), and (C)
regions in the (log $B$,log $\delta$) plane with magnetic field $B \simeq
0.05$ G, and Doppler factor $\delta \simeq 15$.

\begin{figure}
\includegraphics[width=9cm]{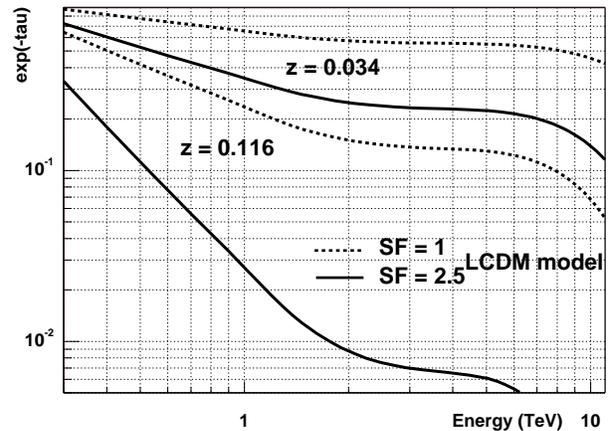}
\caption{Intergalactic absorption factor for 
Mkn~501 ($z$=0.034) and PKS 2155-304 ($z$=0.116)
calculated  for  the LCDM model assuming two different scaling factors,  
$ SF=1$, and $ SF=2.5$.
}\label{fig_abs2}
\end{figure}

If the SSC model is confirmed in future studies,
this would be also a strong evidence 
for  the high CIB at NIR as is shown in Fig.~\ref{fig_new_CIRB},
and used for the reconstruction of the source spectrum of 
the April~16,~1997 flare in Fig.~\ref{fig_sed}. 
Such a high flux of CIB makes significantly steeper ($\Delta \alpha \simeq 0.5$)
the spectrum of  sub-TeV radiation even from a relatively nearby source like 
Mkn~501.  Obviously, the impact to the \gr~emission from farther sources
could be much stronger. In Fig.~\ref{fig_abs2}, the intergalactic 
absorption  factor, $\exp(-\tau_{\gamma \gamma}$)  is shown for Mkn~501 and for the 
BL~Lac object PKS~2155-304 ($z$=0.116), which  recently has been   
reported by the Durham group as a \gr~emitter above 300~GeV 
(Chadwick ~et~al. 1999). It is seen that for a high CIB flux with the scaling factor 
$ SF=2.5$ the \gr~spectrum should suffer 
dramatic steepening ($\Delta \alpha \simeq 2$), before reaching
the observer.  Thus, even for a quite flat source spectrum, {\it e.g.} with photon
index 1.5, we should expect very steep $\gamma$-ray spectrum from this source,
(${\rm d} N/{\rm d}E \propto E^{-3.5}$).
Interestingly, at energies above 2~TeV the absorption becomes almost 
energy independent, therefore the observed spectrum 
essentially repeats the spectral shape of the source spectrum. However, 
the detection of $\geq 2 \, \rm TeV$  radiation of \grs~from  PKS~2155-304
would be very difficult because of  strong,  by a factor of 
100,  suppression of  \gr~fluxes in that energy regime.  

\section{Summary} 

The {accurate spectrometric observations of the 
$\gamma$-ray spectrum of Mkn~501 during the remarkable April~16,~1997 flare 
by the low-threshold  imaging  atmospheric Cherenkov telescope CAT in the
energy region from 300~GeV to 10~TeV is used for
extraction of important information about the CIB at wavelengths
$\geq$ 0.4~\mic. The interpretation of the spectrum of sub-TeV \grs, 
together with simultaneously obtained X-ray data of BeppoSAX,  requires,
within the  one-zone \mbox{SSC} model,  rather high NIR background 
at a level close to  $20 \, \rm nW \, m^{-2} \, sr^{-1}$ at 1~\mic.  
Such high flux of CIB implies an essential distortion of the 
shape of the initial (source) spectrum of $\gamma$-rays from Mkn~501
not only at multi-TeV, but also at sub-TeV energies. 
The `reconstructed' intrinsic $\gamma$-ray spectrum shows a
distinct  peak in the Spectral Energy Distribution 
around 2~TeV with a flux by a factor of 3 higher than   
the measured flux.  Moreover,  the energy spectrum of gamma radiation 
from both sides of the peak has power-law behavior with 
spectral indices $\alpha \simeq 0.5$ below 2~TeV, and
$\alpha \simeq 1.5$ above 2~TeV, which  perfectly  agrees with 
predictions of the \mbox{SSC} model.  We have shown that the intergalactic 
absorption has  non-negligible impact on  the construction
of self-consistent \mbox{SSC} parameters.    

And finally, we argue that  the CAT \gr~data
alone allow rather robust upper limits on the CIB,
$\lambda F_\lambda \leq 60 \ \rm nW \, m^{-2} \, sr^{-1}$ at 1~\mic,
taking into account  that for any reasonable scenario of \gr~production 
the differential intrinsic spectrum of \grs~hardly could be flatter than 
${\rm d} N/{\rm d} E \propto E^{-1}$.

\begin{acknowledgements}
We thank the referee,  E. Dwek  for his critical comments which 
help us to improve significantly the paper. 
We are grateful to  E. Pian and H. Krawczynski who provided us
with  numerical values of BeppoSAX and HEGRA fluxes, as well as
to J. Primack and J. Bullock for  sending us the results of their numerical 
modeling of the CIB. We  thank A. Barrau for fruitful discussions and F. Tavecchio for his help.
F.A.A. thanks the LPNHE-Jussieu for kind hospitality during his stay
at Universit\'e Paris~VII. 
\end{acknowledgements}

\end{document}